\documentclass[11pt,a4paper,final]{article}
\usepackage[pdftex]{graphicx} 
\usepackage[centertags]{amsmath}
\usepackage{amssymb} 
\usepackage{mathrsfs}
\usepackage{gensymb}
\usepackage{color}
\usepackage[a4paper]{geometry}
\usepackage[square,numbers]{natbib} 
\newcommand*{\doi}[1]{\href{http://dx.doi.org/#1}{doi: #1}}
\usepackage{setspace}

\usepackage[pdftitle={An asymptotic approximation to the free-surface
    elevation around a cylinder in long waves}, 
pdfauthor={Michael T. Morris-Thomas}, 
pdfkeywords={water-waves, perturbation theory, diffraction,
  wave-structure interaction},
colorlinks=true,citecolor=red]{hyperref}
\newcommand{\order}{\mathcal{O}}
\newcommand{\eps}{\varepsilon}
\newcommand{\rh}{\hat r}
\newcommand{\zh}{\hat z}
\newcommand{\dip}{\partial}
\newcommand{\MG}{\mathrm{G}}
\begin{document}

\title{\textbf{An asymptotic approximation to the free-surface
    elevation around a cylinder in long waves}} 
\author{Michael T. Morris-Thomas\\
  \small{School of Mechanical Engineering, The University of Western Australia},\\
  \small{35 Stirling Highway, Crawley, Western Australia, 6009.\footnote{E-mail: \texttt{michael.morristhomas@gmail.com}}}}
\date{}%
\maketitle%
%\linespread{1.2}%

% -----------------------------------------------------------------------

\begin{abstract}
  Strong nonlinear effects are known to contribute to the wave run-up
  caused when a progressive wave impinges on a vertical surface
  piercing cylinder. The magnitude of the wave run-up is largely
  dependent on the coupling of the cylinder slenderness, $ka$, and wave
  steepness, $kA$, parameters. This present work proposes an
  analytical solution to the free-surface elevation around a circular
  cylinder in plane progressive waves. It is assumed throughout that
  the horizontal extent of the cylinder is much smaller than the
  incident wavelength and of the same order of magnitude as the incident
  wave amplitude. A perturbation expansion of the velocity potential
  and free-surface boundary condition is invoked and solved to
  third-order in terms of $ka$ and $kA$. The validity of this approach
  is investigated through a comparison with canonical second-order
  diffraction theory and existing experimental results. We find that
  for small $ka$, the long wavelength theory is valid up to $kA
  \approx 0.16-0.2$ on the up wave side of the cylinder. However, this
  domain is significantly reduced to $kA<0.06$ when an arbitrary
  position around the cylinder is considered. An important feature of
  this work is an improved account of the first-harmonic of the
  free-surface elevation over linear diffraction theory.
\end{abstract}

\onehalfspacing

\section{Introduction} 

When a monochromatic progressive wave of fundamental frequency
$\omega$ interacts with a body penetrating the free-surface,
energy is transferred to the harmonics of the carrier wave and the
vertical displacement of the surrounding free-surface is enhanced.
The elevation of the free-surface in contact with the body is
known as the wave run-up, denoted $R$, and is measured from the
mean water level.

We consider a vertical circular cylinder in deep water waves
whereby the incident wave is scattered in all directions along the
horizontal plane of the free-surface. The extent of this
scattering is described by the scattering parameter $ka$ -- where
$k$ and $a$ denote the wave number and cylinder radius
respectively. Provided the wave steepness is small and the
cylinder non-absorbing, the transfer of wave energy implies that
the wave run-up is somewhat less than $2A$. If on the other hand,
the incident wave steepness is sufficiently large, but bounded by
the wave breaking limit, a vertical jet of water at the body
surface will form. This is an extreme form of wave run-up.

%%%%

Linear scattering of water waves by a fixed vertical
cylinder was first considered by \citet[]{Havelock1940} and
extended to second-order by \citet[]{Hunt1981}. In both theories, the
nonlinear free-surface boundary condition --- 
\begin{equation}\label{eq:exact_fsbc}
\phi_{tt} + g\phi_z  =  - 2\nabla \phi  \cdot \nabla \phi _t  -
\tfrac{1} {2}\nabla \phi  \cdot \nabla \left( {\nabla \phi }
\right)^2, \qquad \text{on $z = \zeta$},
\end{equation}
where $\phi$ denotes the velocity potential and $\zeta$ the vertical
free-surface displacement --- is treated by a Stokes expansion in
terms of $kA$ and are consistent to $\order(A)$ and $\order(A^2)$
respectively. The problem of wave scattering by a cylinder has
attracted attention through two alternative treatments of the
nonlinear free-surface boundary condition to $\order(A^3)$ in the work
of \citet*[]{Faltinsen1995} and \citet{Malenica1995}. Of the two
schemes, both assume incident waves of small steepness $kA$, however,
the infinite water depth theory of \citet[]{Faltinsen1995} (hereafter
referred to as `FNV') is restricted to $ka \ll 1$, while that of
\citet[]{Malenica1995} is based on classical Stokes perturbation
expansion and is valid for arbitrary wavelength and water depth. These
two works were stimulated by observations of the excitation of
resonant modes in offshore structures exhibiting high
natural-frequencies by higher-harmonic nonlinear wave loads. Due the
restriction of $ka \ll 1$, the work of FNV is strictly a long
wavelength (LWL) theory.

In realistic ocean environments comprising long wavelengths, the
incident wave amplitude can be of comparable magnitude to the
horizontal extent of a partially immersed cylinder. Consequently, the
free-surface Keulegan-Carpenter number, $K_c=\pi A/a$, is of
$\order(1)$. This clearly violates the Stokes perturbation expansion
of the free-surface boundary condition which implicitly restricts $K_c
\ll 1$. With this in mind, FNV reconsiders the governing relative
length scales to the problem in a consistent fashion such that $kA =
\order(\varepsilon)$, $ka = \order(\varepsilon)$ and $A/a = \order(1)$
--- where $\eps$ is a perturbation quantity of order $\ll 1$. Whilst
the restriction placed on $ka$ essentially implies that the incident
wavelength is asymptotically large when compared to the cylinder
radius, it also implies that the incident wave amplitude is of the
same order of magnitude as the cylinder radius.

In computing wave forces, this LWL approach has been
generalised by \citet[]{Newman1996a} for the case of unidirectional
irregular waves, whereby the wavelength of each spectral component is
assumed large when compared to the cylinder radius. A further
generalisation by \citet{Faltinsen1999}, employing a Lewis conformal
mapping, accounts for the the third-harmonic LWL force associated with
square cylinders comprising rounded edges. 

While for the wave run-up, \citet[]{Newman1995} compared an FNV
second-order free-surface elevation prediction with second-order
diffraction computations. This demonstrated that the two methods are
in reasonable agreement for long wavelengths.  Moreover, experimental
evidence provided by \citet[]{MMT04b} demonstrates that intermediate to
long wavelengths are critical for the wave run-up, particularly when
the amplitudes of these waves are of comparable magnitude to the
cylinder radius. In this regime, third-harmonic components of
$\order(A^3)$ become important, accounting for up to 10\% of the
overall wave run-up. Furthermore, nonlinear effects operating at the
first-harmonic have been clearly observed for increasing values of
$kA$ \citep[]{MMT04b}. Classical linear and second-order diffraction
theories do not account for these nonlinearities. The goal of this
present work is to illustrate that these nonlinearities appear and can
be partially accounted for in the long wavelength regime.

In this present work the LWL theory of FNV is extended to account
for the free-surface elevation correct to $\order(\varepsilon^3)$
for monochromatic progressive waves of amplitude $A$ incident upon
a fixed vertical surface piercing cylinder of radius $a$ in deep
water. Following a somewhat different path than that considered in
FNV, we treat the free-surface boundary condition via two
perturbation schemes corresponding to an expansion in terms of
$ka$ and $kA$. The analytical results are compared to measured
wave run-up values and second-order diffraction calculations.

\section{Problem Definition}

\begin{figure}[t]
\centering
\includegraphics[scale=1.0]{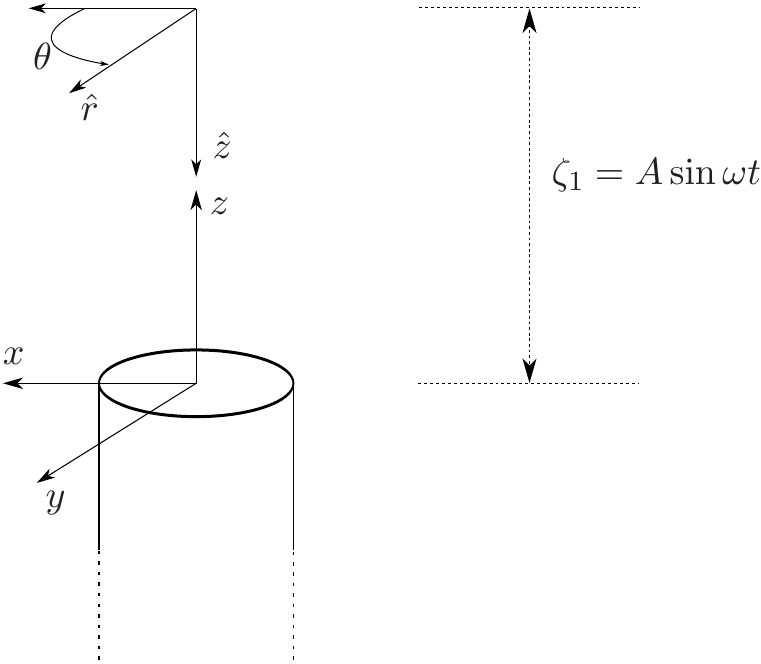}
\caption{Schematic of the coordinate system adopted for long
wavelength theory.} \label{fig:diagram}
\end{figure}

A fluid domain of infinite horizontal extent is considered. The fluid
is assumed ideal and of infinite depth but bounded by
Equation~\eqref{eq:exact_fsbc} at the free-surface. A circular
cylinder of radius $a$ is partially immersed in the fluid. The
coordinate system adopted for the problem is defined in
Figure~\ref{fig:diagram} where $(x, y) = (r \cos \theta, r \sin
\theta)$.  The boundary value problem for the velocity potential
$\phi(r, \theta, z, t)$ is defined:
\begin{align}
\nabla^2 \phi &= 0,& &\text{for $r \in [a,\infty)\cup \theta \in [0,2\pi] \cup z\in[\zeta,-\infty)$},\label{eq:field}\\
\phi_r &= 0,& &\text{at $r = a$},\label{eq:bbc}\\
|\nabla \phi| &\rightarrow 0,& &\text{as $r \rightarrow \infty$ and $z
  \rightarrow -\infty$,}\label{eq:rc}
\end{align}
along with the free-surface boundary condition \eqref{eq:exact_fsbc}
which must be satisfied at the instantaneous free-surface position
$z=\zeta(r,\theta,t)$.

To treat Equation \eqref{eq:exact_fsbc}, it is convenient to express
$\phi$ in terms of a power series with coefficients
$\{\varepsilon^n;\, n=1, \,2, \,3,\, \ldots\}$ such that $\varepsilon
\ll 1$. In canonical water wave diffraction theory one usually selects
$\eps=kA$ --- the wave steepness. All horizontal dimensions are then
scaled according to $ka$, the diffraction parameter whereby
$a=\order(1)$. These two assumptions implicitly imply that $A/a \ll
1$.

Observations in large waves relative to the horizontal dimensions of a
surface piercing body suggest that $A/a \ll 1$ will not always the
case. Therefore, we impose the alternative restriction that $ka =
\order(\eps)$ and, by consequence, $A/a$ is now of $\order(1)$
\citep{Faltinsen1995}. In large waves, this is clearly the case because
typical wavelengths are always much larger than both $a$ and
$A$. These two assumptions, $kA \ll 1$ and $ka \ll 1$, are essential
and form the basis for what is to follow.

To treat the boundary value problem, we formally select the following
perturbation parameters:
\begin{equation}
  \varepsilon_1 = kA, \quad \text{and} \quad \varepsilon_2 = ka,
\end{equation}
which are both of $\order(\eps)$ --- here, and elsewhere $\eps^{m+n}
\equiv \eps_1^m \eps_2^n$. These represent a parameter and a
coordinate expansion respectively. Similar to canonical diffraction
theory we now define a power series expansion for $\phi$. However,
given that we must also take into account $\varepsilon_2$, the
expansion assumes a slightly different form and is written 
\begin{equation}\label{eq:PhiPowerSeries}
\phi = \varepsilon_1 \phi_1 + \varepsilon_1
\varepsilon_2 \phi_2 + \varepsilon_1 \varepsilon_2^2 \phi_3
+ \varepsilon_1^2 \varepsilon_2 \psi_2 + \varepsilon_1^3
\psi_3 +  \order(\eps^4),
\end{equation}
where $\phi_n$ are linear potentials in terms of the wave steepness
and $\psi_2$ and $\psi_3$ are of second- and third-order
respectively. Similarly for the free-surface elevation, we write the
following expansion
\begin{equation}\label{eq:zeta_series}
  \zeta = \zeta_1 + \zeta_2 + \zeta_3 + \order(\eps^4),
\end{equation}
where $\{\zeta_n;\,n=1,2,3\}$ are defined
\begin{subequations}
  \begin{align}
    \label{eq:zeta1_series}
    \order(\eps):&& \qquad
    \zeta_1 =& \varepsilon_1 \eta_{1,0},
    \qquad \\
    \label{eq:zeta2_series}
    \order(\eps^2):&& \qquad
    \zeta_2 =& \varepsilon_1 \varepsilon_2 \eta_{1,1} + 
              \varepsilon_1^2 \eta_{2,0},
    \quad \\          
    \label{eq:zeta3_series}
    \order(\eps^3):&& \qquad
    \zeta_3 =& \varepsilon_1 \varepsilon_2^2 \eta_{1,2} + 
              \varepsilon_1^2 \varepsilon_2 \eta_{2,1} + 
              \varepsilon_1^3 \eta_{3,0} +
              \frac{\varepsilon_1^4}{\varepsilon_2} \eta_{4,-1}.
    \qquad          
    \end{align}
\end{subequations}
The expression for $\zeta_3$ is particularly interesting because it
contains a term that is proportional to $\eps_1^4 / \eps_2$. In
canonical water wave diffraction theory, this term would presumably
appear at fourth-order in the wave steepness. However, in the dual
expansion considered here, it is strictly of third-order. In the
following sections we solve for $\phi_n$, $\psi_2$ and $\psi_3$
and write a more explicit expression for $\zeta$ the free-surface elevation. 

\section{Linear Potentials}\label{sec:LWLLinear}

We first consider $\{\phi_n;\,n=1,2,3\}$. According to the theory of
Stokes, a plane progressive wave of fundamental circular frequency
$\omega$ propagating from $x=-\infty$ can be expressed by the well
known potential
\begin{equation}\label{eq:incident_wave_xyz}
  \phi(x,y,z,t) = \Re \Bigl\{ \frac{gA}{\omega} e^{kz} e^{i(\omega
    t-kx)}\Bigr\}+\order(A^4), 
\end{equation}
where the dispersion relation correct to third-order in wave steepness
is defined $k_0 = k(1 + k^2 A^2)$ where $k_0 = \omega^2/g$
\citep[][Art.~250]{Lamb1932}. When a wave described by
Equation~\eqref{eq:incident_wave_xyz} interacts with a vertical
cylinder, centred at the origin of our coordinate system
(see Figure~\ref{fig:diagram}), the classical linear diffraction potential
that describes the fluid motion is written
\citep[][Eq.~17]{Havelock1940}:
\begin{equation}\label{eq:LDP}
\phi^{(1)} =  \Re \biggl\{ \frac{g A}{\omega} e^{( kz + i \omega
t)} \sum \limits_{m = 0}^\infty \epsilon_m i^{ -m} C_m(kr) \cos m
\theta \biggr\},
\end{equation}
where $\epsilon_m$ denotes the Neumann symbol defined by $\epsilon
_0=1$ and $\epsilon _m=2$ for $m>0$ and 
\begin{equation}\label{eq:SeriesCylinderFns}
C_m(kr) = J_m(kr) - \frac{J_m^{'}(ka)}{H^{(2)'}_m(ka)}
H_m^{(2)}(kr),
\end{equation}
where $J_m$ and $H_m$ denote Bessel and Hankel functions of order $m$
respectively and the primes denotes differentiation with respect to
the argument. Equation~\eqref{eq:LDP} satisfies the boundary
conditions \eqref{eq:field}--\eqref{eq:rc} and the homogeneous form of
Equation~\eqref{eq:exact_fsbc} prescribed on the plane $z=0$. When
both $ka$ and $kr$ are small, Equation \eqref{eq:LDP} can be expanded
into the following approximate form \citep[Eq.~72]{Lighthill1979}:
\begin{equation}\label{eq:phi1_approx}
\phi^{(1)} = \Re \Bigl\{\frac{g A}{\omega} e^{( kz + i \omega t)}
(C_0(kr) - 2 i C_1(kr)\cos \theta - 2 C_2(kr)\cos 2\theta)\Bigr\},
\end{equation}
where the first three coefficients from
Equation~\eqref{eq:SeriesCylinderFns} are
\begin{equation}\label{eq:C0}
  C_0(kr) = 1 + \tfrac{1}{2} (ka)^2 (\ln \tfrac{1}{2} kr + \gamma  + \pi
i/2) - \tfrac{1} {4}(kr)^2 + \ldots,
\end{equation}
\begin{equation}\label{eq:C1}
  C_1(kr) = \frac{k}{2}\biggl( r + \frac{a^2}{r} \biggr) + \ldots,
\end{equation}
and
\begin{equation}\label{eq:C2}
  C_2(kr) = \frac{k}{8}\biggl( r^2 + \frac{a^4}{r^2} \biggr) + \ldots,
\end{equation}
where $\gamma\approx 0.57722$ denotes the Euler-Mascheroni
constant. Upon considering our perturbation parameters, we can see
that Equation \eqref{eq:phi1_approx} is correct to order $\eps_1
\eps_2^2$ and contains contributions of order $\eps_1$ and $\eps_1
\eps_2$. Therefore, we can then equate like terms from Equation
\eqref{eq:phi1_approx} with those of the series expansion
\eqref{eq:PhiPowerSeries} and recover the following expressions
\begin{subequations}
  \label{eq:varphi}
  \begin{align}
    \label{eq:varphi1}
    &\order(\eps_1): && 
    \phi_1=\frac{\omega}{k^2} e^{kz} \cos \omega t,
     \\
    \label{eq:varphi2}
    &\order(\eps_1 \eps_2): &&
    \phi_2=\frac{\omega}{k^2} e^{kz} \sin \omega t \biggl(
               \frac{r}{a} + \frac{a}{r} \biggr) \cos \theta,
    \\
    \label{eq:varphi3}
    &\order(\eps_1 \eps_2^2): &&
    \phi_3=\frac{\omega}{k^2} e^{kz + i \omega t} \biggl[
       \frac{1}{2} \biggl( \ln\tfrac{1}{2} kr + \gamma  + \frac{\pi i}{2}
       \biggr) - \frac{1}{4} \biggl( \frac{r^2}{a^2} - \cos 2\theta
       \biggl(\frac{r^2}{a^2} + \frac{a^2}{r^2}\biggr) \biggr) \biggr].
    \end{align}
\end{subequations}
The expansion~\eqref{eq:phi1_approx}, and expressions
$\{\phi_n;\,n=1,\,2,\,3\}$, are valid for $kr=\order(\eps_2)$ and, in
terms of the wave steepness, purely linear. Moreover, one may also
notice that but virtue of $\phi_1$, the linear vertical free-surface
displacement on the plane $z=0$ is simply
\begin{equation}
\zeta_1 = \eps_1 \eta_{1,0} 
        =  -\eps_1 \, \frac{1}{g} \, \dip_t \phi_1 
        = A \sin \omega t. 
\end{equation}
Equations~(\ref{eq:varphi1}--\ref{eq:varphi3}) provide the first three
terms of the potential $\phi$. Consequently, what remains is for us to
determine the nonlinear potentials $\psi_2$ and $\psi_3$ to complete
the problem to $\order(\eps^3)$.

\section{Nonlinear Potentials}\label{sec:LWLNonLinear}

In solving for $\psi_2$ and $\psi_3$, we must first transfer the
free-surface boundary condition \eqref{eq:exact_fsbc} to the moving
plane $\zeta_1 = A\sin \omega t$. This is the essential feature of the
approach adopted by \citet[]{Faltinsen1995} and we adopt it
here. The reasoning behind it is that the radial gradients on the
right hand side of Equation~\eqref{eq:exact_fsbc} are amplified by our
choice of coordinate expansion $\eps_2=ka$. Consequently, the leading
order contribution from simply substituting $\{\phi_n; \,
n=1,\,2,\,3\}$ into the RHS of Equation~\eqref{eq:exact_fsbc} about
the plane $z=0$ is of $\order(\eps^2)$. This is not what is
required. To overcome this apparent disparate length scale, we can
affect the transfer about the plane $z=\zeta_1$ thereby recovering, on
the RHS of Equation~\eqref{eq:exact_fsbc}, the leading order
contribution of $\order(\eps^3)$ as required. More specifically, this
leading order contribution is in fact of $\order(\eps_1^2\eps_2)$
which implies that this expansion is valid only in the region of
$r=\order(a)$. With this mind, we formally adopt the normalised
coordinates $(\rh, \theta, \zh)$ for the nonlinear potentials:
\begin{equation}
\hat r(r) = \frac{r}{a}, \qquad \text{and} \qquad \hat z(z,t) = \frac{-z +
\zeta_1}{a}.
\end{equation}
With the aid of Expansion~\eqref{eq:PhiPowerSeries}, the transfer of
Equation~\eqref{eq:exact_fsbc} to the plane $\zeta_1 = A\sin \omega t$
is affected by employing a Taylor series expansion about the moving
plane $\zh=0$. Equating like terms to $\order(\eps^3)$, provides the
following two boundary conditions
\begin{alignat}{3}\label{eq:FSBC1}
  & \text{$\order(\eps_1^2 \eps_2)$:}
  \qquad \qquad  \frac{g}{k} \, \dip_{\zh} \psi_2 =& 2
  \Bigr( \dip_{\rh} \phi_{2} \, \dip_{\rh t} \phi_{2} + 
         \frac{1}{\rh^2} \, \dip_{\theta} \phi_{2} \, \dip_{\theta t} \phi_{2} +
         \dip_{\zh} \phi_{1} \, \dip_{\zh t} \phi_{1} \Bigr), \qquad \qquad
\\
\label{eq:FSBC2}
  & \text{$\order(\eps_1^3)$:}
  \qquad \qquad  \frac{g}{k^3} \, \dip_{\zh}\psi_3 =&
  \frac{1}{2} 
  \Bigr( 
    \dip_{\rh} \phi_{2} \, \dip_{\rh} +
    \frac{1}{\rh^2} \, \dip_{\theta} \phi_{2} \,  \dip_{\theta}
  \Bigr) 
  \Bigl( (\dip_{\rh} \phi_{2})^2 + (\dip_{\theta} \phi_{2})^2 \Bigr),
  \qquad \qquad
\end{alignat}
on $\zh=0$. Physically, Equations~\eqref{eq:FSBC1}~and~\eqref{eq:FSBC2}
represent a vertical velocity distribution imposed on an
harmonically oscillating free-surface. By substituting $\phi_1$
and $\phi_2$ into Equations~\eqref{eq:FSBC1} and \eqref{eq:FSBC2} more
explicit expressions result, whence
\begin{equation} \label{eq:FSBC1_explicit}
\dip_{\zh} \psi_2 = \frac{\omega}{k^2} \sin 2 \omega t \biggl(\frac{1}{\rh^4}
- \frac{2}{\rh^2} \cos 2 \theta \biggr),
\end{equation}
and
\begin{equation} \label{eq:FSBC2_explicit} 
\dip_{\zh} \psi_3 = 
2 \frac{\omega}{k^2} \sin^3 \omega t \biggl( 
\cos \theta \Bigl(\frac{1}{\rh^7} - \frac{2}{\rh^5} \Bigr) + 
 \frac{1}{\rh^3} \cos 3 \theta  \biggr).
\end{equation}
Apart from the change of notation, the linear combination of
Expressions~\eqref{eq:FSBC1_explicit} and \eqref{eq:FSBC2_explicit},
the form of $\varepsilon_1^2 \varepsilon_2 \psi_2 + \varepsilon_1^3
\psi_3$, agrees with that of \citet[Eq.~3.10]{Faltinsen1995}.

%%%%%%%%%%%%%%%%%%%%%%%%%%%%%%%%%%%%%%%%%%%%%%%%%%%%%%%%%%%%%%%%%%%%%%%%%%
\subsection{The Boundary Value Problem}\label{sec:LWLBVP}

The form of the free-surface boundary conditions for $\psi_2$ and
$\psi_3$ suggest implicit azimuthal symmetry and oscillatory time
dependence. Consequently, this prompts solutions for $\psi_2$ and
$\psi_3$ in the following forms:
\begin{equation}\label{eq:psiI}
\psi_2 = \frac{\omega}{k^2}\sin 2\omega t \left( \alpha_0(\rh,\zh) +
\alpha_2(\rh,\zh) \cos 2 \theta \right),
\end{equation}
and
\begin{equation}\label{eq:psiII}
\psi_3 = \frac{\omega}{k^2}\sin^3 \omega t \left(
\alpha_1(\rh,\zh)\cos \theta + \alpha_3(\rh,\zh) \cos 3 \theta \right).
\end{equation}
where the unknown set of coefficients $\{\alpha_m(\rh,\zh); \,
m=0,\,1,\,2,\,3\}$ are required and must be chosen to satisfy
Equations~\eqref{eq:FSBC1_explicit} and
\eqref{eq:FSBC2_explicit}. Formally, four two-dimensional boundary
value problems for $\alpha_m(\rh,\zh)$ are defined as follows:
\begin{align}
\nabla ^2 \alpha_m (\rh,\zh) &= 0, & &\text{$\rh\in[1,\infty)\, \cup \, \zh\in[0,\infty)$,}\label{eq:alpha_FE}\\
\dip_{\rh} \alpha_{m} &= 0,& & \text{on $\rh = 1$},\label{eq:alpha_BBC}\\
|\nabla \alpha_m| &\rightarrow 0,& & \text{as $\rh$ and $\zh
\rightarrow \infty$},\label{eq:alpha_rc}\\
\dip_{\zh} \alpha_{m} &= f_m(\rh),& & \text{on $\zh=0$},\label{eq:alpha_FSBC}
\end{align}
where $f_m(\rh)$ is defined
\begin{equation}\label{eq:FNV free surface forcing}
\{1/\rh^4, -4/\rh^5 + 2/\rh^7, -2/\rh^2, 2/\rh^3 \}.
\end{equation}
For these class of problems, a generalised Fourier-Bessel integral
transform is appropriate \citep[page~161]{Lebedev1965}:
\begin{equation}\label{eq:weber_integral}
f(r) = \int_0^{\infty} \frac{\varphi_\lambda(r) \lambda \;
d\lambda} {|H_{\nu}^{'}(\lambda a)|^2} \int_{a}^{\infty} f(\rho)
\varphi_\lambda(\rho)\rho \; d\rho \qquad a<r<\infty,
\end{equation}
where $H_\nu$ denotes a Hankel function of the first kind and
$\varphi_\lambda(r)$ involves the linear combination
\begin{equation}\label{eq:cylinder_function}
\varphi_\lambda(r) = Y_{\nu}^{'}(\lambda a) J_{\nu}(\lambda r) -
J_{\nu}^{'}(\lambda a) Y_{\nu}(\lambda r)
\end{equation}
such that $\varphi_{\lambda}(a) = 2 \pi /a$ and $\varphi_{\lambda}'(a)
= 0$ thereby satisfying the requirements of the Neumann condition at
the cylinder surface. Provided $\sqrt{r}f(r)$ and $\sqrt{r}f'(r)
\rightarrow 0$ as $r \rightarrow \infty$, the integral transform
\eqref{eq:weber_integral} is valid and the radiation condition, for
$\rh \rightarrow \infty$, is then implicitly satisfied.

In solving the boundary value problem, the integral transform is
applied by multiplying the field equation \eqref{eq:alpha_FE} by
$\rh \varphi_\lambda(\rh)$ and integrating from $1$ to $\infty$.
Taking into account the behaviour of the various functions as $\rh
\rightarrow \infty$, the following ordinary
differential equation results
\begin{equation}\label{eq:alpha_ODE}
\dip_{\zh \zh} \hat \alpha_m (\lambda,\zh) - \lambda^2
\hat \alpha_m = \frac{2}{\pi} \dip_{\rh} \alpha_m(1,\zh),
\end{equation}
with the following transformed free-surface condition \eqref{eq:alpha_FSBC}:  
\begin{equation}\label{eq:weber tranform of f}
    \dip_{\zh} \hat \alpha_{m} = \int_1^{\infty} f_m(\rh) \,
     \rh \, \varphi_{\lambda}(\rh) \; d\rh
\end{equation}
on $\zh=0$ for each integer $m$. The particular solution of
Equation~\eqref{eq:alpha_ODE} that satisfies
Equation~\eqref{eq:weber tranform of f}, and by implication the
boundary value problem \eqref{eq:alpha_FE} -- \eqref{eq:alpha_rc}, is
then
\begin{equation}
\label{eq:alpha_hat}
\hat{\alpha}_m(\lambda, \zh) = -\frac{e^{-\lambda \zh}}{\lambda}
\int_1^\infty f_m(\rh) \varphi_{\lambda}(\rh) \rh \; d\rh.
\end{equation}
%
%%%%%%%%%%%%%%%%%%%%%%%%%%%%%%%%%%%%%%%%%%%%%%%%%%%%%%%%%%%%%%

To employ the inversion formula \eqref{eq:weber_integral}, we must
first obtain more explicit expressions for the integrals shown in
Equation~\eqref{eq:alpha_hat}. First, we consider those involving
$\hat{\alpha}_2$ and $\hat{\alpha}_3$. These can be reduced quite
easily by making the substitution $\xi = \lambda \rh$, and applying
the appropriate recurrence relations for cylinder functions
\citep[Eq.~9.1.27~and~Eq.~9.1.30]{Abramowitz1965}, arriving at the
following results:
\begin{equation}\label{eq:alpha2hat}
\hat{\alpha}_2(\lambda,\zh) = \frac{e^{-\lambda \zh}}{\lambda} \big(
Y_1(\lambda) J_3(\lambda) - J_1(\lambda) Y_3(\lambda)\big),
\end{equation}
\noindent and
\begin{equation}\label{eq:alpha3hat}
\hat{\alpha}_3(\lambda,\zh) = -\frac{e^{-\lambda \zh}}{\lambda} \big(
Y_2(\lambda) J_4(\lambda) - J_2(\lambda) Y_4(\lambda)\big).
\end{equation}
The nature of the integrands for $\hat{\alpha}_0$ and $\hat{\alpha}_1$
are somewhat more troublesome and do not permit a reduction by cylinder
recurrence relations. Consequently, we employ the following integral
relation \citep[Eq.~5]{Steinig1972}
\begin{equation}\label{eq:BesselInt1}
\int x^\mu  \mathscr{C}_\nu(x) \; dx = x \bigl( {(\mu  + \nu  -
1)\mathscr{C}_\nu (x)S_{\mu - 1,\nu  - 1} - \mathscr{C}_{\nu - 1}
(x)S_{\mu ,\nu }} \bigl),
\end{equation}
where $S_{\mu,\nu}(x)$ denotes the Lommel function of the second-kind
of argument $z$ and $\mathscr{C}_\nu$ is any cylinder function of
order $\nu$. In solving for $\hat{\alpha}_0$, we utilise the
substitution $\xi=\lambda \rh$, as before, and apply the integral
relation provided by Equation \eqref{eq:BesselInt1}. After making use
of the Wronskian, and noting that the series $S_{\mu,\nu}\sim
x^{\mu-1}$ for $\xi \rightarrow \infty$, thus ensuring that there is
no contribution from the upper limit of integration, we have
\begin{equation}\label{eq:alpha0hat}
  \hat{\alpha}_0(\lambda,\zh) = - \frac{8}{\pi}\lambda^2 e^{-\lambda \zh}
  S_{-4,1}(\lambda).
\end{equation}
An expression for the function $\hat{\alpha}_1$ can be developed
in a similar manner but with added complexity due to the form of
$f_1(\rh)$. With some manipulation, the following expression is emitted
\begin{equation}\label{eq:alpha1hat}
  \hat{\alpha}_1 = \lambda^4 e^{-\lambda \zh} \biggl(
  \frac{32}{\pi \lambda} S_{-5,0} - \frac{24}{\pi} \lambda
  S_{-7,0}\, + \bigl(2 S_{-4,1} - \lambda^2 S_{-6,1} \bigr)
  \bigl(J_2 Y_0 - Y_2 J_0 \bigr)\biggr),
\end{equation}
where the argument $\lambda$ of the Bessel and Lommel functions is
implied but omitted for brevity.

Finally, expressions for $\{\alpha_m(\rh,\zh);\,m=0,\,1,\,2,\,3\}$ are
determined by inversion in accordance with
Equation~\eqref{eq:weber_integral}, whence
\begin{equation}
\label{eq:alpha_soln}
  \alpha_m(\rh,\zh) = \int_0^{\infty} 
  \hat{\alpha}_m(\lambda,\zh)
  \varphi_{\lambda}(\rh)
  \frac{\lambda \; d\lambda} {|H_{m}^{'}(\lambda)|^2}.
\end{equation}
For $\alpha_0$ and $\alpha_1$ we represent the series
$S_{\mu,\nu}(\lambda)$ using an appropriate hypergeometric form
\citep[Eq.~8.4.27.3]{Prudnikov1990}:
\begin{equation}\label{eq:Suv}
    S_{\mu,\nu} \bigl(2 \sqrt{x} \bigr) = 
    \frac{2^{\mu-1}}{c(\mu,\nu)}
    \, \MG^{3\,1}_{1\,3} \bigg( x \bigg|
    \begin{array}{lll}
    (\mu+1)/2  &        &        \\
    (\mu+1)/2, & \nu/2, & -\nu/2
    \end{array}\bigg),
  \end{equation}
with
\begin{equation}\label{eq:Suvcoeff}
  c(\mu,\nu) = \Gamma \bigl( (1-\mu - \nu)/2 \bigr) \; 
  \Gamma \bigl( (1-\mu + \nu)/2 \bigr),
\end{equation}
where $\MG$ and $\Gamma$ denote Meijer G- and Gamma functions
respectively. Although such a generalised representation of
$S_{\mu,\nu}$ does pose some numerical difficulties, this is our only
recourse due to the fact that the addition of $\mu+\nu$ amounts to odd
negative integer values for each appearance of a Lommel function in
$\alpha_0$ and $\alpha_1$. Consequently, more convenient expressions
for $S_{\mu,\nu}$ \citep[\S\,3]{Steinig1972} are not possible.

The integrands \eqref{eq:alpha_soln} for $m=0$, 1, 2 and 3 are
sufficiently smooth and rapidly approach zero as $\lambda \rightarrow
\infty$ suggesting that its principal contribution arises in the
vicinity of the lower limit of integration. The numerical integration
was conducted in Sage~\citep{sage} which provides a wrapper for the
GNU Scientific Library's \citep{GSL08} adaptive Gaussian-Kronrod
quadrature algorithm QAG\footnote{Numerical integration algorithms in
  the GNU Scientific Library are based on QUADPACK, see
  \texttt{http://nines.cs.kuleuven.be/software/QUADPACK/}.}. The
Meijer G-function \eqref{eq:Suv} was tackled with the mpmath
\citep{mpmath} Python library for arbitrary-precision floating-point
arithmetic.

\begin{figure}[p]
  \centering
  \includegraphics[scale=1]{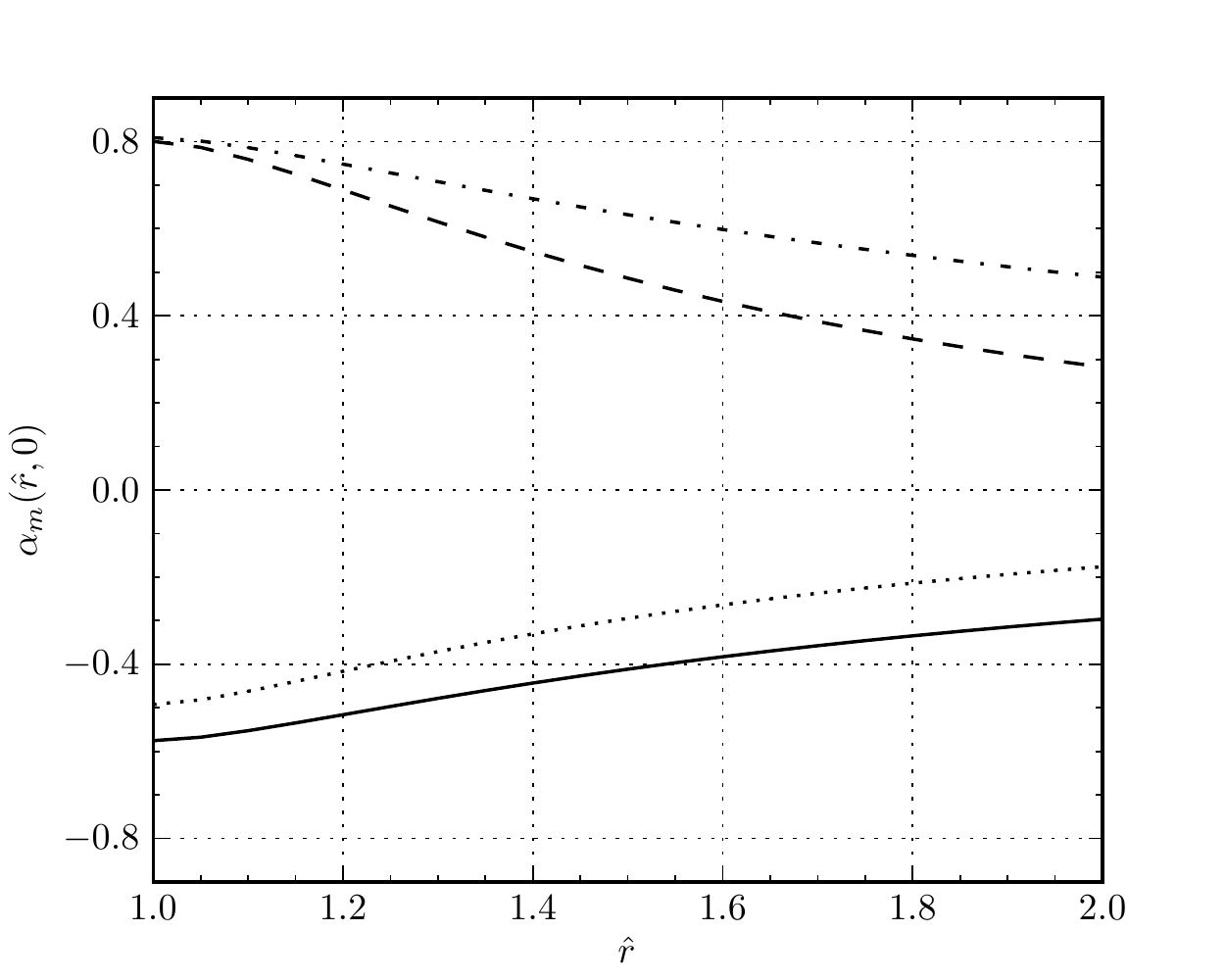}
  \caption{The coefficients $\alpha_m(\rh,0)$ as a
    function of the scaled radial distance for:\,
    $m=0$ (\,-----\;);
    $m=1$ (\,-- -- --\;);
    $m=2$ (\,-- $\cdot$ --\;) ;
    and $m=3$ (\,$\cdots \cdot$\;).} \label{fig:alpha_versus_R}
\end{figure}

Results of the numerical integration for $\{\alpha_m;
m=0,1,2,3\}$ on $\zh=0$ with $1 \le \rh < 2$ are illustrated in
Figure~\ref{fig:alpha_versus_R}. Although each coefficient
monotonically decreases for increasing $\rh$, this attenuation is
rather slow and suggests a significant nonlinear forcing in the
neighbourhood of the cylinder.

\section{The Free-surface Elevation}\label{sec:LWLFSE}

We now turn our attention to the free-surface elevation
\eqref{eq:zeta_series} and determine the coefficients $\eta_{m,n}$ of
the expansion. In its exact form, the free surface elevation follows
from the Bernoulli equation
\begin{equation}\label{eq:exfe}
  \zeta(r,\theta,t)  =  - \frac{1} {g} \Bigl( \phi _t  + \tfrac{1}
  {2} \nabla \phi \cdot \nabla \phi \Bigr),
\end{equation}
where $z-\zeta=0$ defines the instantaneous free-surface position. As
is customary, we transfer Equation~\eqref{eq:exfe} to the plane $z=0$ by
employing a Taylor series expansion. In terms of our normalised
coordinate system for the nonlinear potentials, this plane corresponds
to $\zeta_1=\zh a$. Incorporating the expansion for the velocity
potential \eqref{eq:PhiPowerSeries} along with the perturbation series
for the free-surface elevation \eqref{eq:zeta_series} in the Bernoulli
equation \eqref{eq:exfe} emits the following coefficients for Expressions \eqref{eq:zeta1_series}--\eqref{eq:zeta3_series}\\
$\order(\eps)$:\\
\begin{equation}\label{eq:eta11}
  \eta_{1,0} =  - \frac{1}{g} \, \dip_t \phi_{1},
\end{equation}
$\order(\eps^2)$:\\
\begin{equation}\label{eq:eta21}
  \eta_{1,1} = -\frac{1}{g} \, \dip_t \phi_{2},
\end{equation}
\begin{equation}\label{eq:eta22}
  \eta_{2,0} = -\frac{1}{g} 
  \biggl( \frac{k^2}{2}
  \Bigl( \phi_{1}^2 + ( \dip_{\rh} \phi_{2})^2 + 
  \frac{1}{\rh^{2}} (\dip_{\theta} \phi_{2})^2 
  \Bigr) -
  \frac{1}{g} (\dip_t \phi_{1})^2 
  \biggr), 
\end{equation}
$\order(\eps^3)$:\\
\begin{equation}\label{eq:eta31}
  \eta_{1,2} = -\frac{1}{g} \, \dip_t \phi_{3},
\end{equation}
\begin{equation}\label{eq:eta32}
  \eta_{2,1} = -\frac{1}{g}
  \biggl( \dip_t \psi_2 - \frac{2k}{g} \, 
  \dip_t \phi_{1} \, \dip_t \phi_{2}  + k^2 
  \Bigl( \phi_{1} \, \phi_{2} + \dip_{\rh} \phi_{2} \,
  \dip_{\rh} \phi_{3} +\frac{1}{\rh^2} 
  \, \dip_{\theta} \phi_{2} \, \dip_{\theta} \phi_{3}  
  \Bigr)
  \biggr),
\end{equation}
\begin{multline}\label{eq:eta33}
  \eta_{3,0} = -\frac{1}{g} \biggl( \dip_t \psi_3 + \dip_{\zh} \psi_2
  \,\omega \,\cos \omega t + k^2 \Bigl( \dip_{\rh} \phi_{2} \,
  \dip_{\rh} \psi_2 + \frac{1}{\rh^2}\, \dip_{\theta} \phi_{2}\,
  \dip_{\theta} \psi_2 - \phi_{1}\, \dip_{\zh} \psi_2
  \Bigr)+\\
  + \frac{3}{2}\frac{k^2}{g^2} \, \dip_t \phi_{1}^3 -
  \frac{1}{2}\frac{k^3}{g} \, \dip_t \phi_{1} \, \phi_{1}^2 -
  \frac{3}{2}\frac{k^3}{g} \Bigl(\dip_t \phi_{1}
  (\dip_{\rh}\phi_{2})^2 + \frac{1}{\rh^2}\, \dip_{t} \phi_{1} \,
  \dip_{\theta} \phi_{2} \Bigr) \biggr),
\end{multline}
\begin{equation}\label{eq:eta34}
  \eta_{4,-1} = -\frac{1}{g} 
  \biggl( \dip_{\zh} \psi_3 \omega \cos \omega t + k^2 
  \Bigl( \dip_{\rh} \phi_{2} \, \dip_{\rh} \psi_3 
  +\frac{1}{\rh^2} \, \dip_{\theta}\phi_{2} 
  \, \dip_{\theta}\psi_3 - \phi_{1} \dip_{\zh} \psi_3 
  \Bigr)
  \biggr).
\end{equation}
More explicit expressions for the free-surface elevation follow by
substituting $\phi_1$, $\phi_2$, $\phi_3$, $\psi_1$ and $\psi_2$
\eqref{eq:varphi1}--\eqref{eq:varphi3}, \eqref{eq:psiI} and \eqref{eq:psiII}
into the above equations for $\eta_{m,n}$.

For the free-surface elevation at the surface of the cylinder we set
$\rh=1$, and to second-order we require $\eta_{1,1}$ and
$\eta_{2,0}$. The linear combination of these in accordance with
Equation~\eqref{eq:zeta2_series} gives
\begin{equation}\label{eq:zeta2_on_R=1}
  \zeta_2 = \frac{1}{2}kA^2 \bigl(\cos 2\theta - \tfrac{1}{2} \bigr)
  - 2kAa \cos \theta \cos\omega t + 
  \frac{1}{2}kA^2 \bigl( \cos 2\theta - \tfrac{1}{2} \bigr)\cos 2\omega t
\end{equation}
which essentially agrees with \citet[Eq.~3.13 with $R=1$ in their
notation]{Faltinsen1995}. Similarly, for the third-order free-surface
elevation $\zeta_3$ at $\rh=1$ we can make use of
Expansion~\eqref{eq:zeta3_series} with
Equations~\eqref{eq:eta31}--\eqref{eq:eta34}. After organising
$\zeta_3$ into its harmonic components, the following compact
expression results
\begin{equation} \label{eq:zeta3_on_R=1} \zeta_3 = \sum_{m=1}^3
  a_{m}(\theta) \sin m \omega t + \sum_{m=0}^4 b_{m}(\theta) \cos m
  \omega t,
\end{equation}
where the odd and even coefficients of the series are defined
\begin{align*}
  a_{1} &= \frac{1}{2} k^2 A a^2 \bigl( \ln \bigl(\tfrac{ka}{2} \bigr)
  + \gamma - \tfrac{1}{2} - \cos 2 \theta \bigr)
  + \frac{1}{4} k^2 A^3 \bigl( 9\cos 2 \theta - 5 \bigr ),\\
  a_{2} &= -\frac{1}{2} k^2 A^2 a
  \bigl(5\cos \theta + \cos 3\theta \bigr), \\
  a_{3} &= \frac{1}{4} k^2 A^3 ( 1 - 3\cos 2\theta ), \intertext{and}
  b_{0} &= -\frac{3 k^2}{8} \frac{A^4}{a} \Bigl(\alpha_1 +
  \bigl(3\alpha_3 - \alpha_1 \bigr)\cos 2\theta -
  3\alpha_3 \cos 4\theta \Bigr),\\
  b_{1} &= \frac{\pi k^2}{4} A a^2 - k^2 A^3 \Bigl( \bigl(
  \tfrac{3}{4}\alpha_1 + \alpha_2 \bigr) \cos \theta +
  \bigl(\tfrac{3}{4}\alpha_3 - \alpha_2 \bigr) \cos 3\theta
  \Bigr),\\
  b_{2} &= -2 k^2 A^2 a \bigl( \alpha_0 + \alpha_2 \cos 2\theta \bigr)
  + \frac{k^2}{2} \frac{A^4}{a} \Bigl( \alpha_1 + \bigl(3\alpha_3 -
  \alpha_1 \bigl) \cos 2 \theta - 3 \alpha_3 \cos 4 \theta
  \Bigr),\\
  b_{3} &= k^2 A^3 \Bigl( \bigl(\alpha_2 + \tfrac{3}{4} \alpha_1
  \bigr) \cos \theta - \bigl(\alpha_2 - \tfrac{3}{4} \alpha_3 \bigr)
  \cos 3\theta
  \Bigr),\\
  b_{4} &= -\frac{k^2}{8} \frac{A^4}{a} \Bigl( \alpha_1 +
  \bigl(3\alpha_3 - \alpha_1 \bigr) \cos 2\theta - 3\alpha_3 \cos
  4\theta \Bigr),
\end{align*}
In the above expressions, it is understood that $\alpha_m$ is defined
on the plane $\zh=1$ and the dispersion relation $\omega^2 /g = k (1 +
k^2A^2)$ holds.

It is interesting to note that while $\eta_{3,0}$ contributes to both
the first- and third-harmonics, $\eta_{4,-1}$ contributes to the
zeroth-, second- and fourth-harmonics. These contributions are
consistent with a conventional Stokes expansion in terms of
$\varepsilon_1$ only. In terms of $\varepsilon_2$,
$\varepsilon_1^3\eta_{3,0}$ is the leading order contribution of a
Stokes expansion to $\order(A^3)$. Moreover, although
$\varepsilon_1^4/\varepsilon_2\eta_{4,-1}$ is of fourth-order in wave
amplitude, it is strictly of $\order(\varepsilon^3)$ in the dual
expansion considered here.

\section{Discussion}

It is important to reiterate that a conventional Stokes expansion,
whereby $A \ll L$, when applied to the free-surface boundary
condition, implicitly restricts $A \ll a$. Ostensibly, this is by
virtue of the fact that $a=\order(L)$. In contrast, the theory
presented here explicitly assumes $A/a=\order(1)$. This is a salient
point and constitutes the underlying difference between the two
schemes. 

Consequently, certain contributions present in both $\psi_2$ and
$\psi_3$ only arise due to the intrinsic assumption $A/a =
\order(1)$. For example, additional first- and second-harmonic
contributions, not present in either $\zeta_1$ or $\zeta_2$, result
from the forcing of both Equations~\eqref{eq:varphi1} and
\eqref{eq:varphi2} at the free-surface.  Some of these additional
contributions would in fact be accounted for in canonical diffraction
theory to $\order(A^2)$ and $\order(A^3)$ while those of
$\order(A^4/a)$ would not. In other words, contributions of
$\order(A^4/a)$ are strictly inconsistent with a Stokes expansion to
third-order wave steepness.

\begin{figure}[p]
\centerline{\includegraphics*[scale=1]{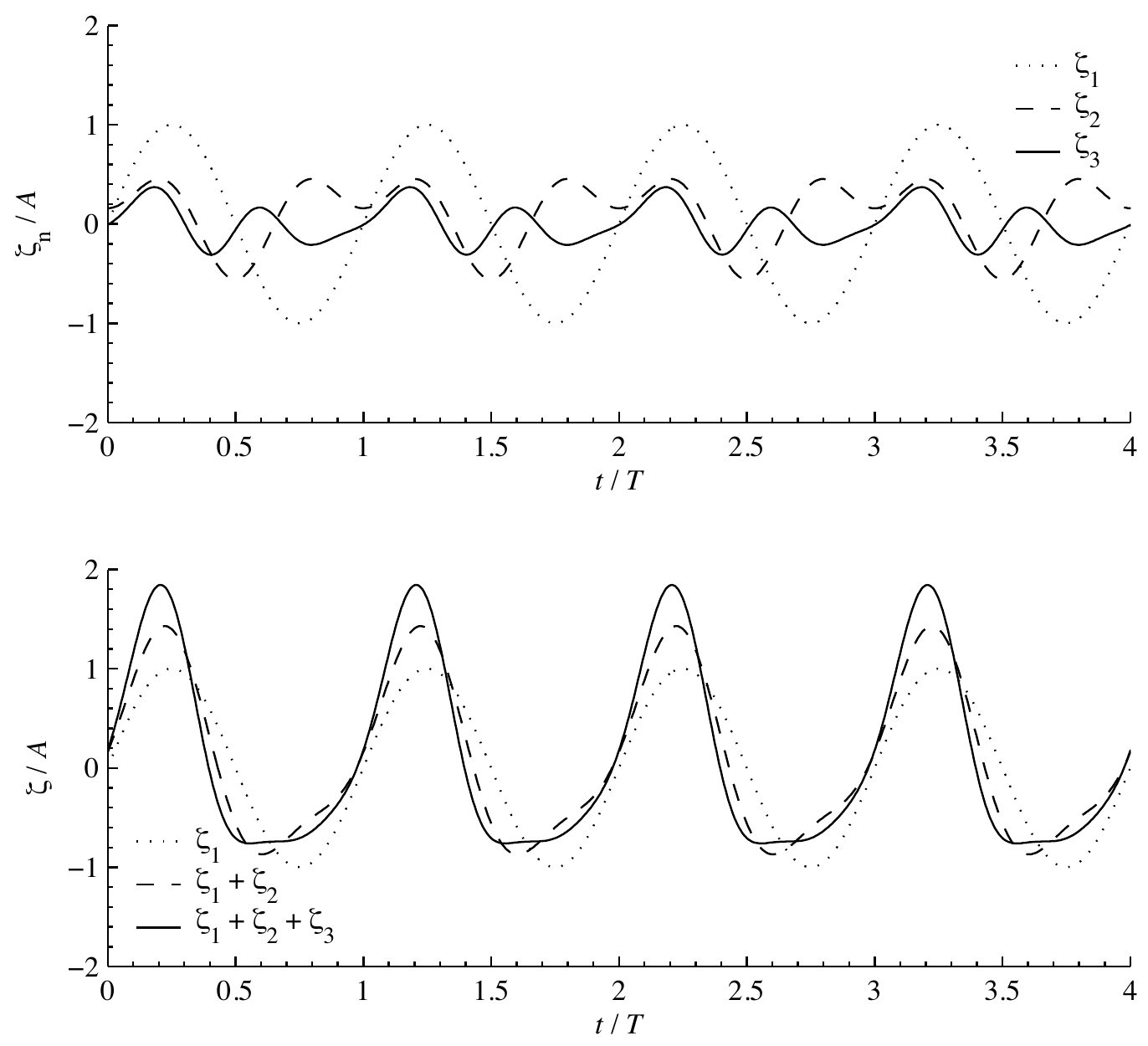}}
\caption{Time series representation of the wave run-up on a vertical
  cylinder at $\theta=\pi$ for $ka=0.208$ and $kA = 0.4$. The top
  figure shows the contributions of each order $\zeta_1$, $\zeta_2$
  and $\zeta_3$. The bottom figure shows the systematic addition of
  each of these components.}
\label{fig:FNV time series}
\end{figure}

To illustrate the wave run-up at $\theta=\pi$, produced by the present
theory, we present a time trace, Figure~\ref{fig:FNV time series}, for
$ka=0.208$ and $kA = 0.4$. Although this wave is in fact approaching
the wave breaking limit in deep water, $kA_{\text{max}} = 0.14\pi$, it
does serve to illustrate third-order contributions arising from
$\zeta_3$. In comparison to $\zeta_1$, second- and third-order
components are approximately $0.37\zeta_1$ and $0.08\zeta_1$
respectively. It is worth noting that $\zeta_3$ emits one time
independent term that does not contribute to the wave run-up at
$\theta=\pi$ due to an equal and opposite forcing at the
free-surface. Consequently, only $\zeta_2$ contributes to the
zeroth-harmonic at $\theta=\pi$ and is purely of $\order(A^2)$ in
origin.

\begin{figure}[p]
  \centerline{\includegraphics*[scale = 1]{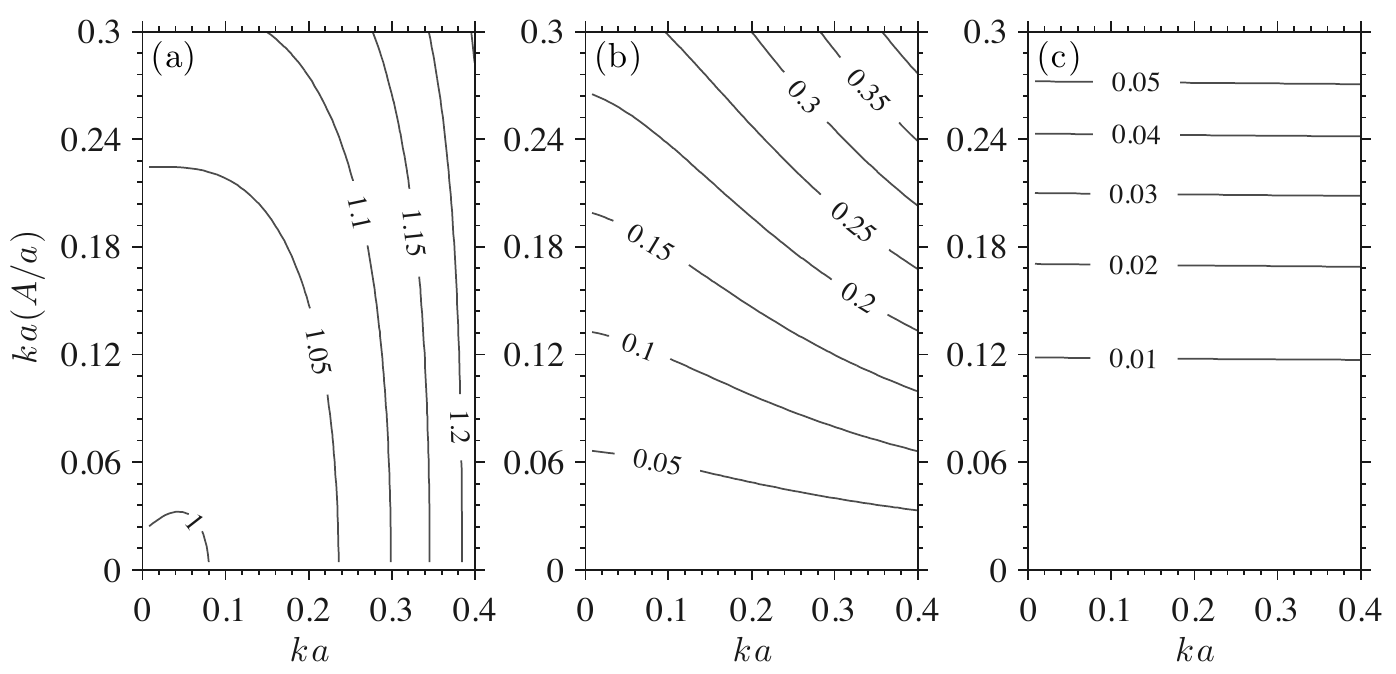}} %
  \caption{Contour plots of the harmonic components of the
    free-surface elevation computed at the position $(r, \theta) = (a,
    \pi)$ plotted against the scattering parameter $ka$ and wave
    steepness $kA = ka(A/a)$: (a) the first-harmonic; (b) the
    second-harmonic; and (c) the third-harmonic.}    
  \label{fig:zetacontours}
\end{figure}

More general results for the free-surface elevation at
$(r,\theta)=(a,\pi)$ are presented in
Figure~\ref{fig:zetacontours}. We observe that the wave run-up is
magnified with both the scattering parameter and the wave
steepness. However, this magnification with wave steepness is
considerably more dramatic and gradually increases with increasing
$ka$. In conventional diffraction theory \citep{Havelock1940}, the wave
run-up at $\theta=\pi$ should approach 2 as $ka \rightarrow
\infty$. In the present theory, however, this is not the case because
it is only valid for asymptotically small $ka$

\begin{figure}[p]
\centerline{\includegraphics*[scale = 1]{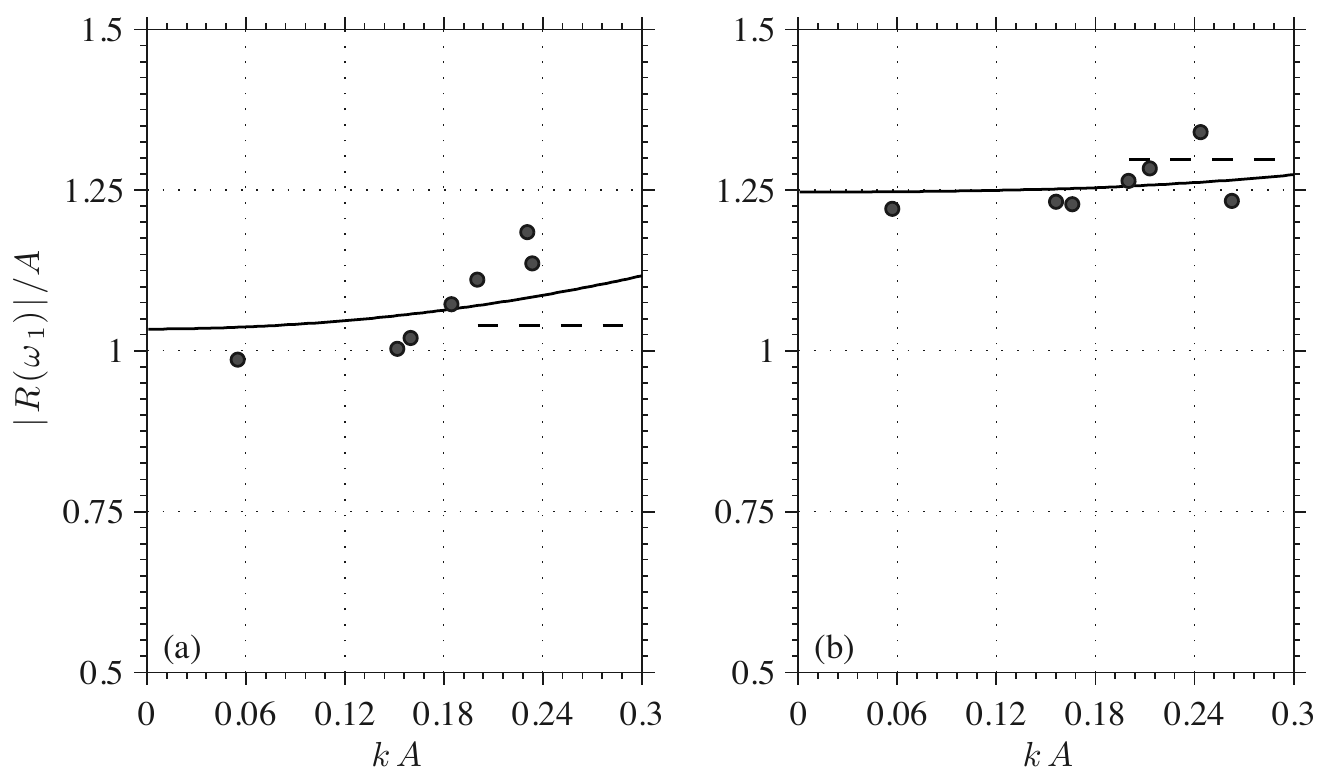}} %
\caption{The normalised modulus of the first-harmonic component of the
  wave run-up $|R(\omega_1)|/A$, evaluated at the incident wave side
  of the cylinder $(r=a,\theta=\pi)$, versus the wave steepness
  $kA$. The two axes correspond to scattering parameters of (a)
  $ka=0.208$ and (b) $ka=0.417$. The data sets are: (\,------\,)
  present third-order LWL computations; (\,-- -- --\,) linear
  diffraction theory; and ({\tiny $\bullet$}) measured values
  \cite{MMT04b}.}
 \label{fig:Plot_zeta1_LWL_kA_1H}
\end{figure}

To investigate the effectiveness of the present theory in predicting
the harmonic components of the wave run-up we first consider the
fundamental frequency. Figure~\ref{fig:Plot_zeta1_LWL_kA_1H} compares
the first-harmonic prediction,
\begin{equation*}
\zeta^{(1)}(1,\pi,t) = \eps_1 \eta_{1,0} + \eps_1 \eps_2
\eta_{1,1}+\chi_{1,1}\sin \omega t + \chi_{2,1}\cos \omega t,
\end{equation*}
with linear diffraction theory \citep{Havelock1940} and measured
first-harmonics of the wave run-up on a circular cylinder
\citep[]{MMT04b} --- the first-harmonic wave run-up is denoted
$R(\omega_1)$. We observe that although the present formulation
under predicts the measured first-harmonic for large $kA$ it appears
to be an improvement over the linear diffraction formulation. This 
observation is particularly evident for values of $ka=0.208$ and
$0.417$. On the other hand, the LWL theory is clearly invalid for
$ka=0.698$. This is ostensibly because the assumption of $ka \ll
1$ has been violated and has therefore compromised the comparison.

\begin{figure}[p]
\centerline{\includegraphics*[scale=1]{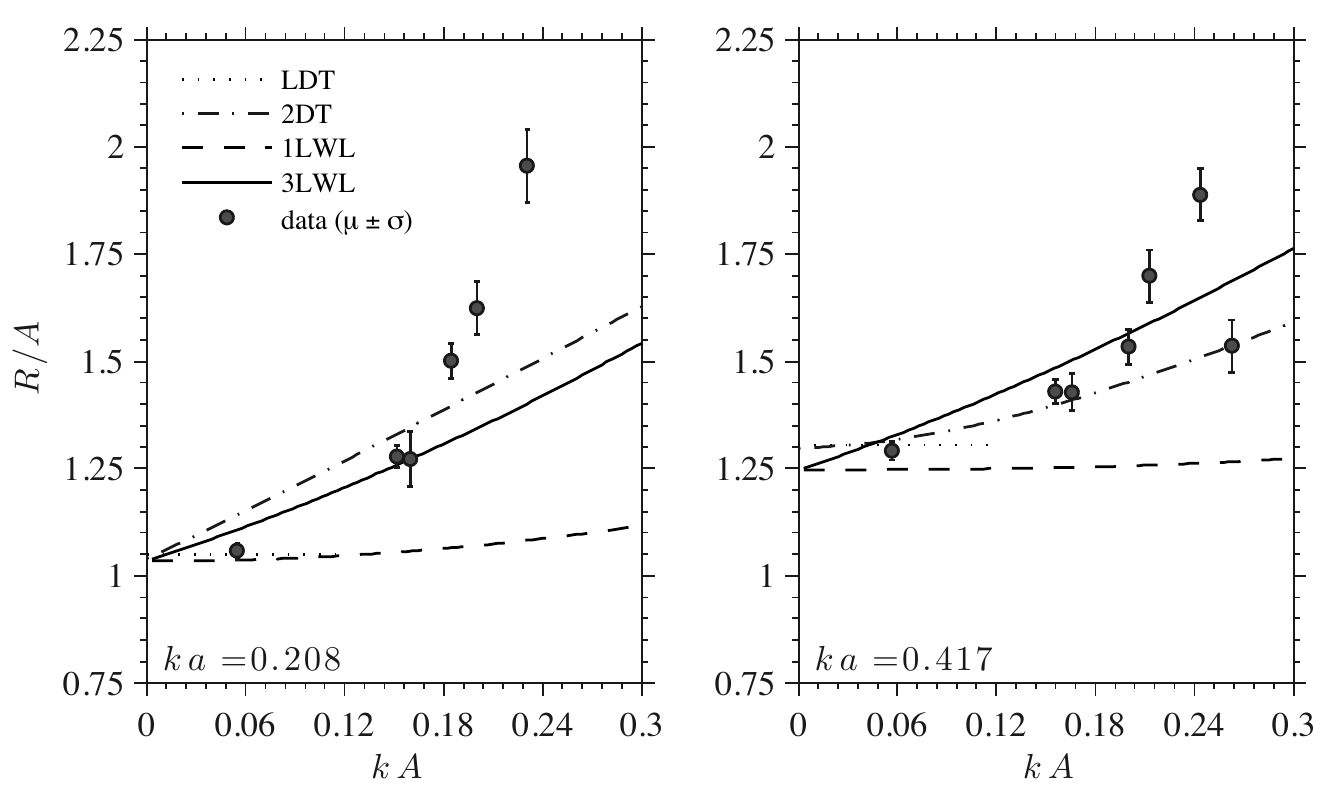}}
\caption{The maximum wave run-up $R/A$ evaluated at $(r=a,\theta=\pi)$
  plotted against the wave steepness $kA$. The present long wavelength
  theory is denoted 1LWL and 3LWL for the first- and third-order
  computations; LDT and 2DT correspond to linear and second-order
  diffraction results respectively. The data ($\mu \pm \sigma$)
  corresponds to measured values of the run-up in plane progressive
  waves \cite{MMT04b}; $\mu$ and $\sigma$ denote the mean and standard
  deviation of the measured $R$ respectively.}
 \label{fig:Rmax_compare}
\end{figure}

We now investigate the validity of the present theory by comparing the
wave run-up to measured values \citep[]{MMT04b}. To accommodate this,
we present Figure~\ref{fig:Rmax_compare} which shows the normalised
wave run-up $R/A$, at the position $\theta=\pi$, versus the infinite
depth wave steepness $k_0A$. Also included are, correct to
second-order in wave steepness, diffraction computations from the
boundary integral equation method program \textsc{Wamit} and linear
diffraction theory \citep{Havelock1940}. Our
first observation is that, for $ka=0.698$, the present theory
distinctly over-predicts the wave run-up. However, this is not
surprising as the condition $ka \ll 1$ is not sufficiently satisfied
and local diffraction effects associated solely with $ka$ are clearly
important here.

In contrast, the present theory performs well for both $ka=0.417$ and
$0.208$ in the domain of $k_0A < 0.15-0.2$ (see
Figure~\ref{fig:Rmax_compare}). Interestingly, both second-order
diffraction computations and the present third-order theory agree for
$ka=0.417$ and, in addition, both compare favourably to the measured
results. However, for $ka=0.208$, the present theory agrees more
favourably with measured values than second-order diffraction
computations. In particular, this observation distinctly holds true
for $k_0A$ less than approximately $0.15$. Moreover, this suggests than
third-order wave steepness nonlinearities are significant in long
waves and in fact more important than nonlinearities associated with
local wave diffraction associated with $ka$.

\begin{figure}[p]
\centerline{\includegraphics[scale=1]{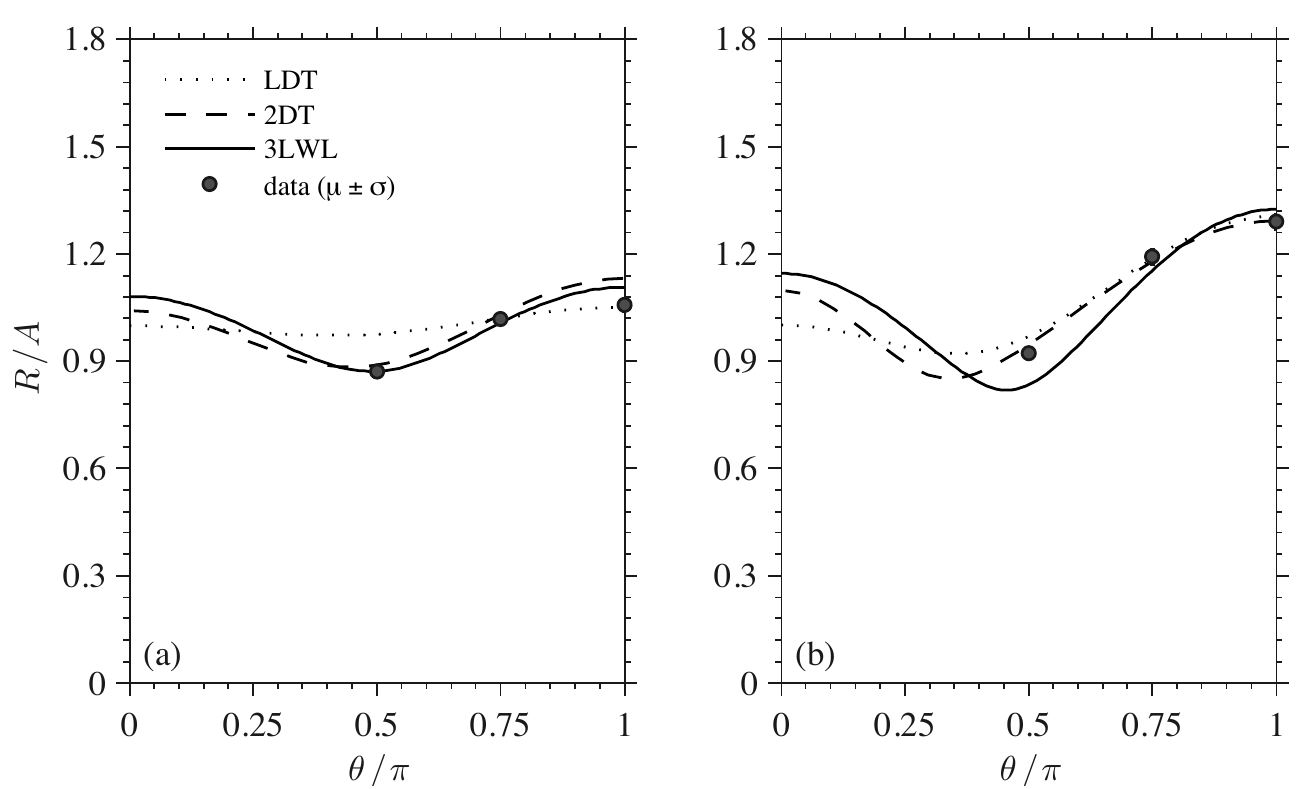}}
\caption{The maximum wave run-up around a truncated vertical
  cylinder ($d/a=2.53$) in plane progressive waves of steepness
  $kA=0.016$ for scattering parameters: $ka=0.208$ (a); and $ka=0.417$
  (b). The variation in the maximum free-surface elevation $R/A$ with
  $\theta/\pi$ is shown for the present third-order long wavelength
  (3LWL) theory, linear (LDT) and second-order (2DT) diffraction
  theories, and measured data \cite[]{MMT04b}.} \label{fig:Figure_theta}
\end{figure}

We now turn our attention to the free-surface elevation around the
boundary of the cylinder. Figure~\ref{fig:Figure_theta} illustrates
two plots of $R(\theta)/A$ versus $\theta/\pi$ for the moderate wave
steepness $kA=0.016$. While the first plot concerns the relatively
small scattering $ka=0.208$, the second concerns a more moderate value
$ka=0.417$. Of the two, the first is indicative of the range of
validity of the present theory. Whereby the incident wavelength is
much larger than the cylinder's diameter.

In all cases we observe a local minimum of $R(\theta)/A$ in the region
of $0 < \theta/\pi < 0.5$ (Figure~\ref{fig:Figure_theta}). Physically,
this results from an increase in fluid momentum as the flow negotiates
the cylinder, hence to conserve energy, the free-surface must decrease
in elevation accordingly. It appears as though this local minimum is a
function of both local diffraction $ka$ and wave steepness
$kA$. Presumably, this minimum shifts closer to $\theta/\pi = 0.5$ in
the limit of $ka \rightarrow 0$ because the cylinder becomes
increasingly transparent to the onset flow. This observation is
consistent with the present theory.

Of the two plots, Figure~\ref{fig:Figure_theta}a demonstrates that the
present theory captures the trend exhibited by the measured data when
local diffraction effects, associated with $ka$, are small. Moreover,
the theory also agrees with second-order diffraction
computations. Also shown, is the improvement by the present theory over
linear diffraction theory (Figure~\ref{fig:Figure_theta}). For more
moderate scattering, $ka=0.417$ for instance, while the present theory
correctly predicts the amplification of the wave run-up at
$\theta=\pi$, it slightly under-predicts the the amplification at
$\theta/\pi=0.5$ -- by approximately 9\% (see
Figure~\ref{fig:Figure_theta}b). This under-prediction is likely caused
by dominant $\cos \theta$ symmetry terms associated with the
second-harmonic contributions in the expression for $\zeta_3$. Despite
this, the present theory does capture the overall trend of the
measured data.

\section{Conclusions}\label{sec:conclusions}

We have described a long wavelength theory, correct to
$\order(\eps^3)$, for the free-surface elevation around a vertical
cylinder in plane progressive waves. It provides an efficient means to
evaluate the wave run-up provided that the intrinsic assumptions of
$A/a = \order(1)$, $ka \ll 1$ and $kA \ll 1$ are observed. Indeed,
after utilising measured results, we have demonstrated that the
present theory performs well provided these assumptions are satisfied.

Salient features of the solution to the free-surface elevation are
nonlinear effects of $\order(A^3)$ -- operating at the first- and
third-harmonics, and contributions of $\order(A^4/a)$ -- operating at
the zeroth-, second-, and fourth-harmonics. For instance, these
nonlinear effects provide an improved account of the first-harmonic
wave run-up over linear diffraction theory when local diffraction
effects associated with $ka$ are small. Moreover, when these local
diffraction effects are small the present theory agrees more
favourably with the overall wave run-up at $\theta=\pi$ than
second-order diffraction theory computations. 

\subsection*{Acknowledgements}
The author acknowledges the financial support of the Australian
Research Council, MARINTEK, and Molly Roberts through the late
Professor Hew Roberts Trust at the University of Western
Australia. The comments of Dr. Grant Keady (UWA), Dr. Krish
Thiagarajan (UWA), and Dr. J\o rgen Ranum Krokstad (Statkraft) are
gratefully acknowledged.

% References

\clearpage

% images

%---------------------------------------------------------------
%---------------------------------------------------------------

% --------------------------------------------------------------

% ------------------------------------------------------------
% Numerical results 
% -------------------------------------------------------------

% ... /wave-run-up/analysis/lwl/time-series
\begin{figure}[p]
  \centerline{\includegraphics*[scale = 1]{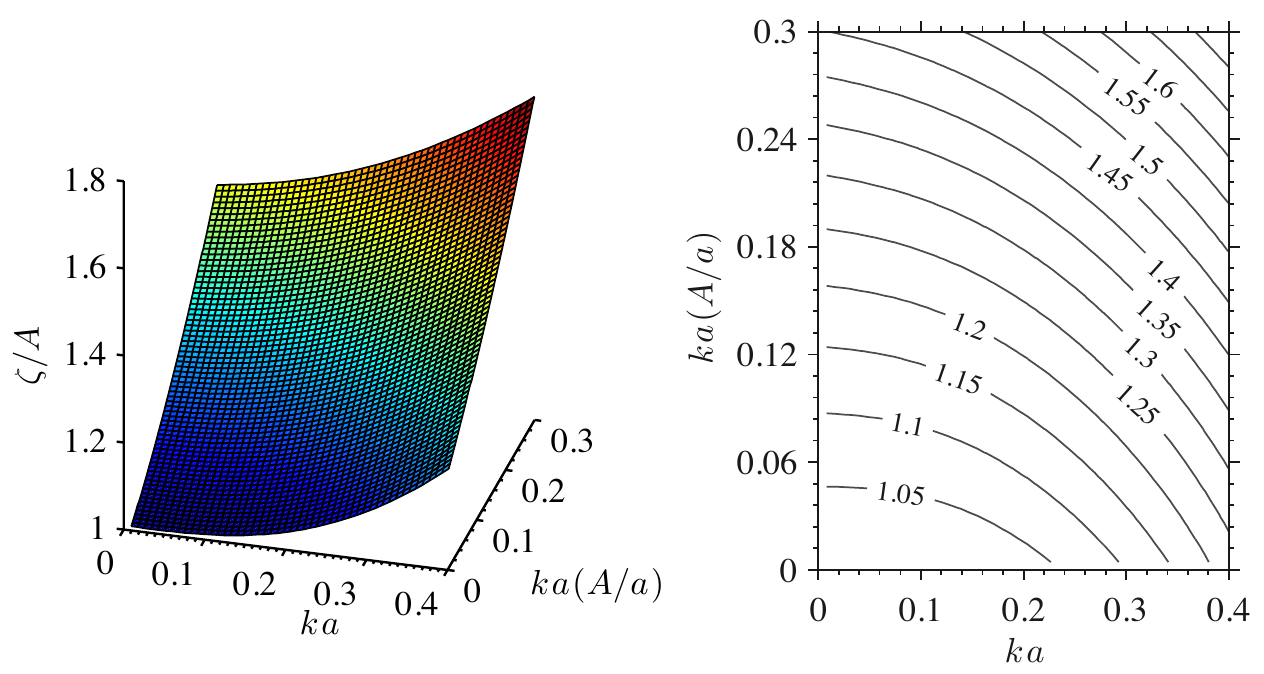}} %
  \caption{Surface and corresponding contour plots of the maximum
    free-surface elevation $\zeta / A$ on the upwave side of the
    cylinder surface from LWL theory. Here, $\zeta / A$ is evaluated
    at the position $(r, \theta) = (a , \pi)$ and plotted against the
    scattering parameter $ka$ and wave steepness $kA = ka(A/a)$.}
  \label{fig:zetamax}
\end{figure}

% ... /wave-run-up/analysis/lwl/time-series

% ------------------------------------------------------------
% Comparison with experiments plots
% -------------------------------------------------------------

% First harmonic plots
% ... /wave-run-up/analysis/lwl/first-harmonic/

% Maximum wave run-up plots
% ... /wave-run-up/analysis/lwl/maximum-runup/

% Variation of the wave run-up with theta plots
% ... /wave-run-up/analysis/lwl/maximum-runup/

%--------------------------------------------------------------

% \alpha_0 = -0.5754128306
% \alpha_1 =  0.7949762183
% \alpha_2 =  0.8089847469
% \alpha_3 = -0.4924778843
\end{document}